\documentclass[chicago]{emulateapj}
\usepackage{apjfonts}
\usepackage{graphicx}

\newcommand{\wcen}{\omega\ {\rm Cen}}
\newcommand{\jk}{J-K_s}
\newcommand{\teff}{T_{\rm eff}}
\newcommand{\logg}{\log{g}}

\slugcomment{Last modified: \today}

\shorttitle{Medium-Resolution Spectroscopy in $\omega$ Centauri}
\shortauthors{An et~al.}

\begin{document}

\title{Medium-Resolution Spectroscopy of Red Giant Branch Stars in $\omega$ Centauri}

\author{Deokkeun An\altaffilmark{1},
Young Sun Lee\altaffilmark{2},
Jae In Jung\altaffilmark{1},
Soo-Chang Rey\altaffilmark{2},
Jaehyon Rhee\altaffilmark{3,6},\\
Jae-Woo Lee\altaffilmark{4},
Young-Wook Lee\altaffilmark{5},
Young Hoon Joe\altaffilmark{5,6}
}

\altaffiltext{1}{Department of Science Education, Ewha Womans University, 52 Ewhayeodae-gil, Seodaemun-gu, Seoul 03760, Korea; deokkeun@ewha.ac.kr}
\altaffiltext{2}{Department of Astronomy and Space Science, Chungnam National University, Daejeon 34134, Korea}
\altaffiltext{3}{Harvard-Smithsonian Center for Astrophysics, 60 Garden Street, MS-09, Cambridge, MA 02138, USA}
\altaffiltext{4}{Department of Physics and Astronomy, Sejong University, 209 Neungdong-ro, Gwangjin-Gu, Seoul, 05006, Korea}
\altaffiltext{5}{Center for Galaxy Evolution Research and Department of Astronomy, Yonsei University, Seoul 03722, Korea}
\altaffiltext{6}{Visiting astronomer, Cerro Tololo Inter-American Observatory, National Optical Astronomy Observatory, which is operated by the Association of Universities for Research in Astronomy (AURA) under a cooperative agreement with the National Science Foundation.}

\begin{abstract}

We present [Fe/H] and [Ca/Fe] of $\sim600$ red giant branch (RGB) members of the globular cluster Omega Centauri ($\wcen$). We collect medium-resolution ($R\sim2000$) spectra using the Blanco 4 m telescope at the Cerro Tololo Inter-American Observatory equipped with Hydra, the fiber-fed multi-object spectrograph. We demonstrate that blending of stellar light in optical fibers severely limits the accuracy of spectroscopic parameters in the crowded central region of the cluster. When photometric temperatures are taken in the spectroscopic analysis, our kinematically selected cluster members, excluding those that are strongly affected by flux from neighboring stars, include relatively fewer stars at intermediate metallicity ([Fe/H]$\sim-1.5$) than seen in the previous high-resolution survey for brighter giants in Johnson \& Pilachowski. As opposed to the trend of increasing [Ca/Fe] with [Fe/H] found by those authors, our [Ca/Fe] estimates, based on \ion{Ca}{2} H \& K measurements, show essentially the same mean [Ca/Fe] for most of the metal-poor and metal-intermediate populations in this cluster, suggesting that mass- or metallicity-dependent SN~II yields may not be necessary in their proposed chemical evolution scenario. Metal-rich cluster members in our sample show a large spread in [Ca/Fe], and do not exhibit a clear bimodal distribution in [Ca/Fe]. We also do not find convincing evidence for a radial metallicity gradient among RGB stars in $\wcen$.

\end{abstract}


\section{Introduction} \label{sec:introduction}

Omega Centauri ($\wcen$) is the most massive globular cluster found in the Milky Way Galaxy. Extensive studies on this cluster have revealed the existence of multiple stellar populations with a spread in metallicity and/or age, leading to a general consensus that $\wcen$ is a remnant nucleus of a tidally stripped dwarf galaxy in the Milky Way \citep[e.g.,][]{norris:95,suntzeff:96,lee:99,smith:00,piotto:05,lee:09}. The majority of stars in the red giant branch (RGB) of the cluster have metallicities peaked at [Fe/H]$\sim-1.8$, and there are extended metal-rich tails to [Fe/H]$\sim-0.5$ \citep[e.g.,][hereafter JP10]{johnson:10}. The most metal-poor and metal-intermediate RGB populations are often dubbed RGB-MP and RGB-MInt, respectively, where RGB-MInt populations are further divided into three sub-groups \citep[RGB-MInt1, RGB-MInt2, and RGB-MInt3;][]{sollima:05}. In addition, \citet{pancino:00} identified an anomalous RGB (hereafter RGB-a) with significantly redder colors than the main RGB populations, making them the most metal-rich population in the cluster.

Large spectroscopic abundance studies are useful ways of dissecting complex chemical evolutions of stars and star formation histories in $\wcen$. To date, JP10 conducted the most extensive high-resolution spectroscopic survey on the cluster. They obtained spectra for $855$ stars brighter than $V=13.5$ in the cluster, and provided detailed elemental abundances for these stars. They found that abundances of heavy $\alpha$-elements ([$\alpha$/Fe]), such as calcium and silicon, show a complex morphology as a function of [Fe/H]. Most notably, their estimates on the fractional abundance of calcium relative to iron ([Ca/Fe]) systematically increase almost by $0.1$~dex from [Ca/Fe]$=+0.26$ at [Fe/H]$\la-1.6$ (RGB-MP) to [Ca/Fe]$=+0.34$ at $-1.6\la {\rm [Fe/H]}\la-1.3$ (RGB-MInt1). They proposed mass- and/or metallicity-dependent SN~II yields as a possible explanation for their observed abundance trends.

In addition, JP10 made a tentative conclusion that RGB-MInt2+3 stars ($-1.3 \la {\rm [Fe/H]}\la-0.9$) show a bimodal [Ca/Fe] distribution and are split in equal proportion into two groups, each of which is peaked at [Ca/Fe]$=+0.25$ and $+0.45$, respectively. Such trends have never been observed toward Galactic globular clusters, but could have profound implications for understanding the formation and evolution of $\wcen$ and studying its relation to the build-up of stellar halos. For example, \citet{nissen:10} argued that a large fraction of halo stars in the Milky Way originated from the progenitor dwarf galaxy of $\wcen$. Stars in the Galactic halo are divided into two spatially overlapping populations, the inner and the outer halos, which have distinct spatial, kinematical, and chemical properties from each other \citep{carollo:07,beers:12,an:13,an:15}. Their argument is based on bimodal distributions of [$\alpha$/Fe] observed in nearby halo stars \citep[see also][]{fernandez:17}, according to which relatively low-$\alpha$ stars belong to the outer halo. However, whether the outer halo stars are linked to the Ca-poor sequence in $\wcen$ remains unclear, primarily due to a limited sample size of metal-rich cluster members.

In this paper, we present the [Fe/H] and [Ca/Fe] of $\sim600$ RGB members of $\wcen$. Our sample is at least $0.7$~mag fainter in $V$ than the JP10 sample, providing an opportunity to independently check the above findings. In addition, the effective temperatures ($\teff$) of these stars are in the regime ($\teff > 4500$~K), where the infrared flux method (IRFM) relations are well-defined. Our sample selection and observation are described in \S~\ref{sec:sample}. Stellar parameters, including [Fe/H] and [Ca/Fe], are presented in \S~\ref{sec:param}. The effects of crowding in the spectroscopy analysis are assessed. Our newly obtained metallicity distribution function (MDF) of $\wcen$ stars and their [Ca/Fe] distributions are inspected in \S~\ref{sec:result}.

\section{Observations and Data Reductions}\label{sec:sample}

\begin{figure}
\centering
\epsscale{1.05}
\plotone{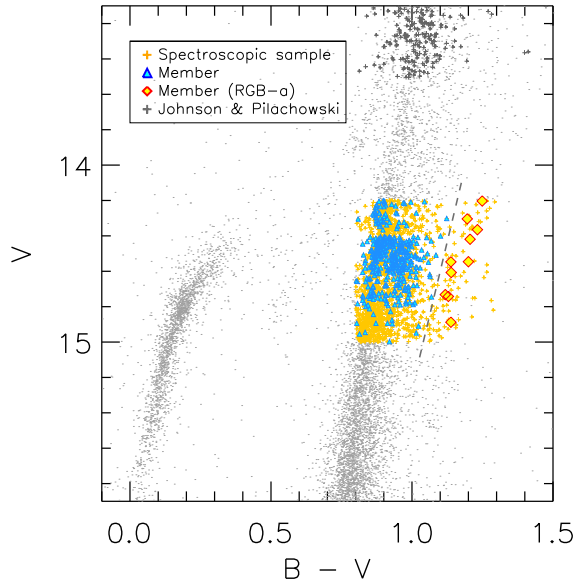}
\caption{$BV$ CMD of $\wcen$ from RLR04. Spectroscopic targets are indicated by an orange plus point. The blue triangles represent kinematically selected cluster members, which also have good spectroscopic parameter estimates without significant contaminations by nearby stars (see the text). Among them, RGB-a members are highlighted by a big red diamond. The sloped dashed line shows a photometric division between RGB-a stars and the rest of the cluster stars. Spectroscopic targets observed in the previous high-resolution survey by JP10 are shown by black crosses.\label{fig:cmd}} \end{figure}

We selected spectroscopic targets among RGB stars in $\wcen$ based on a $BV$ color-magnitude diagram (CMD) of \citet[][hereafter RLR04]{rey:04}. The gray points in Figure~\ref{fig:cmd} represent all photometric objects observed in the $\sim40\times40$~arcmin$^2$ field centered at the cluster in RLR04. We applied a color-magnitude cut with $14.2 < V < 15.0$ and $0.8 < B-V < 1.3$. Stars in this relatively narrow magnitude bin are as bright as or fainter than the average RR Lyrae brightness in the cluster. Our sample is at least $0.7$~mag fainter in $V$ than the faintest objects observed in JP10 (black crosses).

We collected medium-resolution spectra using the Blanco 4 m telescope equipped with Hydra at the Cerro Tololo Inter-American Observatory (CTIO). The Hydra is a fiber-fed multi-object spectrograph with $138$ large optical fibers, each of which has a $\sim300\,\micron$ diameter or a $\sim2\arcsec$ aperture on a focal plane. We observed $\wcen$ in $11$ different fiber configurations on UT 2005 May 27 and 28. The sky was mostly clear, with intermittent scattered sky conditions. The lunar illumination was $\sim70\%$, but the sky background was relatively low due to a large angular separation of the Moon. Observation of each configuration was repeated $2$--$4$ times, with $10$--$15$~min long integration times. The BG39 filter and KPGL1 grating provided $R\equiv \lambda/\Delta\lambda\sim2000$, covering $\sim3700$\, \AA -- $6065$\, \AA. We selected and observed an unbiased, random subset of stars ($N=968$), which is approximately one-third of all photometric objects in the above color-magnitude range. Most of them ($N=807$) were observed three times; $90$ stars were observed four times, and $71$ stars were observed twice.

We followed standard IRAF\footnote{IRAF (Image Reduction and Analysis Facility) is distributed by the National Optical Astronomy Observatory, which is operated by the Association of Universities for Research in Astronomy, Inc., under cooperative agreement with the National Science Foundation.} procedures to pre-process two-dimensional images of the Hydra spectra. Specifically, we used the CCDPROC task to subtract a linear fit to the overscan region, trim image edges, and subtract the averaged nightly bias frame from the image frames. We then extracted spectra from the pre-processed images using the DOHYDRA task. We began by removing cosmic rays in each frame by setting a proper threshold in DOHYDRA. We used projector flats (exposures of a quartz lamp illuminating the fibers), which have been taken for each combination of fiber configuration and telescope pointing, to trace individual fiber profiles and set extraction apertures. We used dome flats (exposures of an illuminated white spot) to remove variations in the pixel-to-pixel sensitivity of the CCD chip and to correct for a difference in the fiber throughput. We utilized HeNeArXe comparison lamp exposures to obtain wavelength solutions, which typically have a root mean square (r.m.s.) scatter of $0.06$\ \AA, or $\sim1/20$ of the wavelength interval per pixel ($1.154$\ \AA). About $20$ fibers were allocated for sky observations in each fiber configuration. We obtained an averaged sky spectrum for each configuration, and subtracted it from all object spectra. Finally, we interpolated the extracted spectra to a common linear wavelength scale.

In this study, we present spectroscopic parameters of $712$ stars (including cluster non-members) for which there exist at least one spectrum, with a signal-to-noise ratio (S/N) equal to or greater than $52$ per resolution elements at $4500 \leq \lambda ({\rm \AA}) \leq 5500$. In Figure~\ref{fig:cmd}, the orange crosses represent these $712$ spectroscopic sample stars. Among them, cluster members with reliable spectroscopic parameter estimates are indicated by blue triangles (see below). We further divided our sample and selected RGB-a members (shown by red diamonds) using a dashed line in Figure~\ref{fig:cmd}. We drew the line by hand, since the RGB-a sequence is well separated from the rest of the RGBs of the cluster on the $BV$ CMD. It is also close to the one adopted in RLR04 to select the reddest RGB in the cluster (the `most metal-rich' [MMR] stars in their terminology).

\section{Stellar Parameters and Cluster Membership}\label{sec:param}

\subsection{[Fe/H], $\teff$, $\logg$, and $v_r$}

We derived the stellar parameters from each of the observed spectra, by searching for the best-matching model spectrum. We adopted and modified one of the spectral matching techniques, dubbed NGS1, in the SEGUE \citep[Sloan Extension for Galactic Understanding and Exploration;][]{yanny:09} Stellar Parameter Pipeline \citep[SSPP;][]{sspp1,sspp2} in the Sloan Digital Sky Survey \citep{sdss}. SSPP has successfully passed a series of tests for validating the accuracy of spectroscopic parameters using high-resolution spectra and Galactic cluster data \citep{sspp3,sspp1,sspp2}. Recently, \citet{kim:16} applied the modified spectral matching technique, like the one used in this work, to high-S/N ($>100$) spectra of bright $G$- and $K$-type dwarfs in the solar neighborhood, after downgrading spectra to $R\sim10,000$, to validate stellar parameters against those based on equivalent-width analysis using MOOG \citep{sneden:73}.

To briefly summarize the spectral matching technique adopted in this study \citep[see also][]{kim:16}, we employed the $\chi^2$ minimization routine MPFIT \citep{markwardt:09} to search the grid of synthetic spectra for the best-fitting set of $\teff$, $\logg$, and [Fe/H] for an observed spectrum. Radial velocity ($v_r$) measurements were followed through cross-correlation with a synthetic spectrum. To generate a grid of synthetic spectra, we used the ATLAS9 model atmospheres based on the new opacity distribution functions \citep{castelli:04}\footnote{Available at http://kurucz.harvard.edu/grids.html}. By linearly interpolating a pre-computed set of the ATLAS9 models, we created a finer grid of model atmospheres, covering $4000$~K $< \teff < 10,000$~K in steps of $250$~K, $0.0 < \logg < 5.0$ in steps of $0.25$~dex, and $-5.0 < {\rm [Fe/H]} < +1.0$ in steps of $0.25$~dex. We used the ATLAS9 {\it synthe} code to generate synthetic spectra in the wavelength range of $3000$--$10,000$\ \AA\ at a resolution of $0.01$\ \AA. For an input to {\it synthe}, we computed [$\alpha$/Fe] for a given [Fe/H], assuming an approximate relation found in the Galaxy: [$\alpha$/Fe]$=+0.4$ at [Fe/H]$\leq-1$, [$\alpha$/Fe]$=0.0$ at [Fe/H]$\geq0.0$, and a linear interpolation at $-1.0 < {\rm [Fe/H]} < 0.0$.

In this work, we used data at $4500 \leq \lambda ({\rm \AA}) \leq 5500$ for matching with the synthetic grid. The region includes H$\beta$, which is a sensitive indicator of $\teff$, and contains a large number of isolated Fe lines and various metallic lines. Since our fitting includes all the observed data within the given wavelength range, our derived metallicity is driven not only by Fe lines, but also by various metallic lines in the spectra. Nevertheless, our [Fe/H] values presented in the following analysis indicate [Fe/H] in the ALAS9 spectral library. To expedite the parameter search, but without a loss of accuracy, we smoothed the model spectrum to $R=1000$ at $5000$\ \AA, and resampled to $1.0$\ \AA-wide linear pixels. We normalized the model spectra using a pseudo-continuum, which was constructed by iteratively rejecting data points that are more than $1\sigma$ below or $4\sigma$ above a fitted polynomial curve. As done for the model spectra, we degraded (to $R=1000$), re-binned (to $1.0$\ \AA\ wide pixels), and normalized the observed spectra.

\begin{figure}
\centering
\includegraphics[scale=0.35]{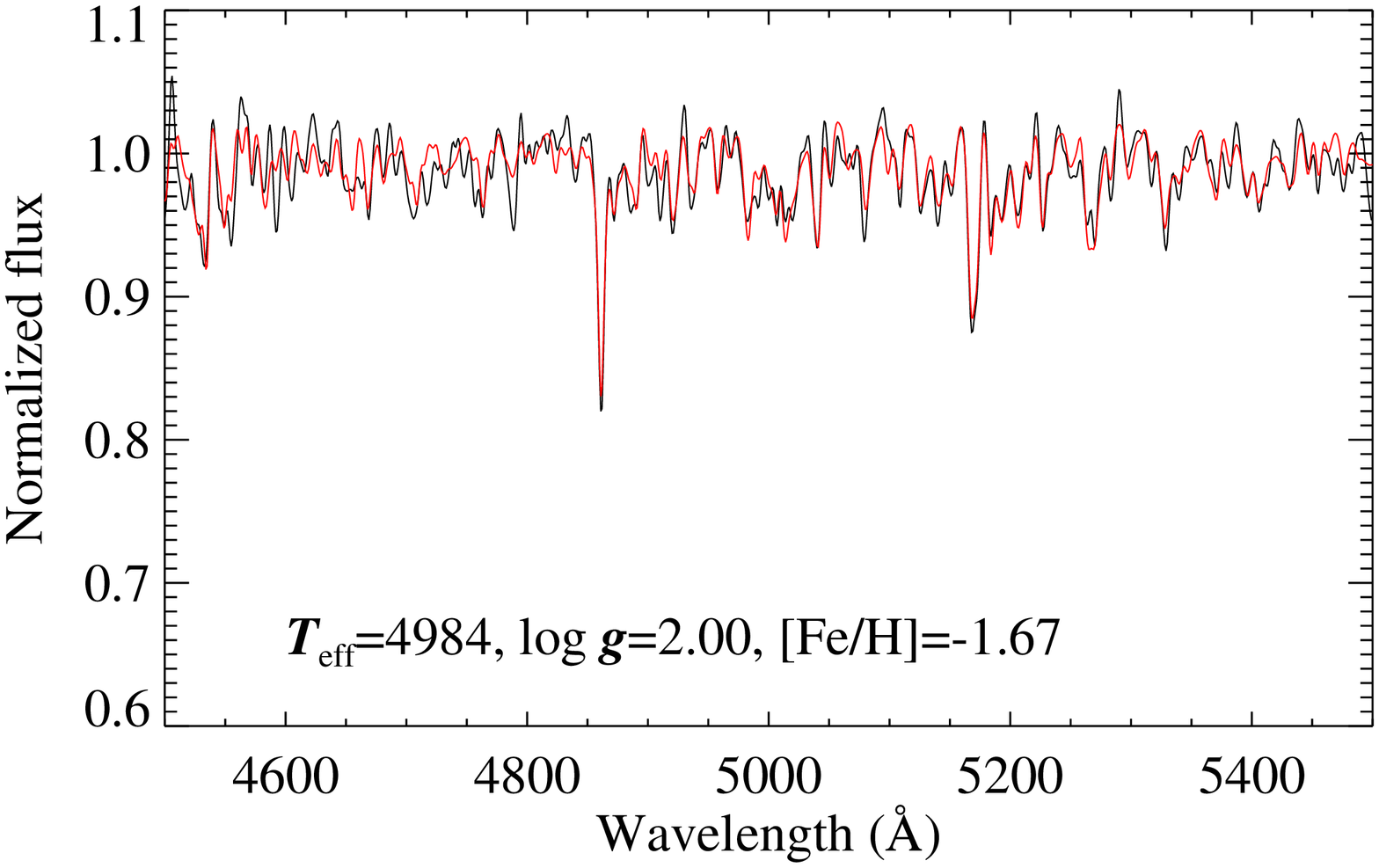}
\includegraphics[scale=0.35]{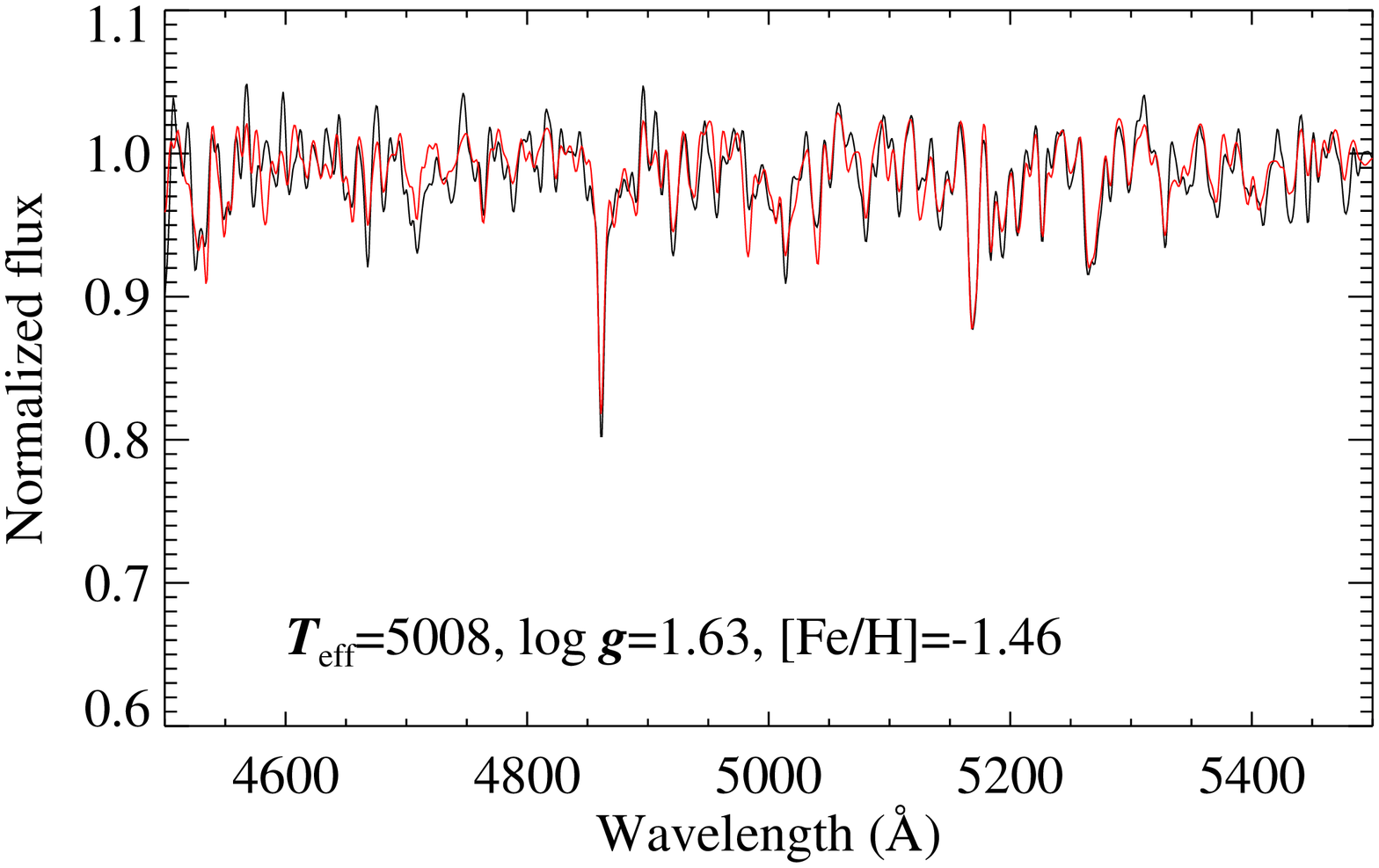}
\includegraphics[scale=0.35]{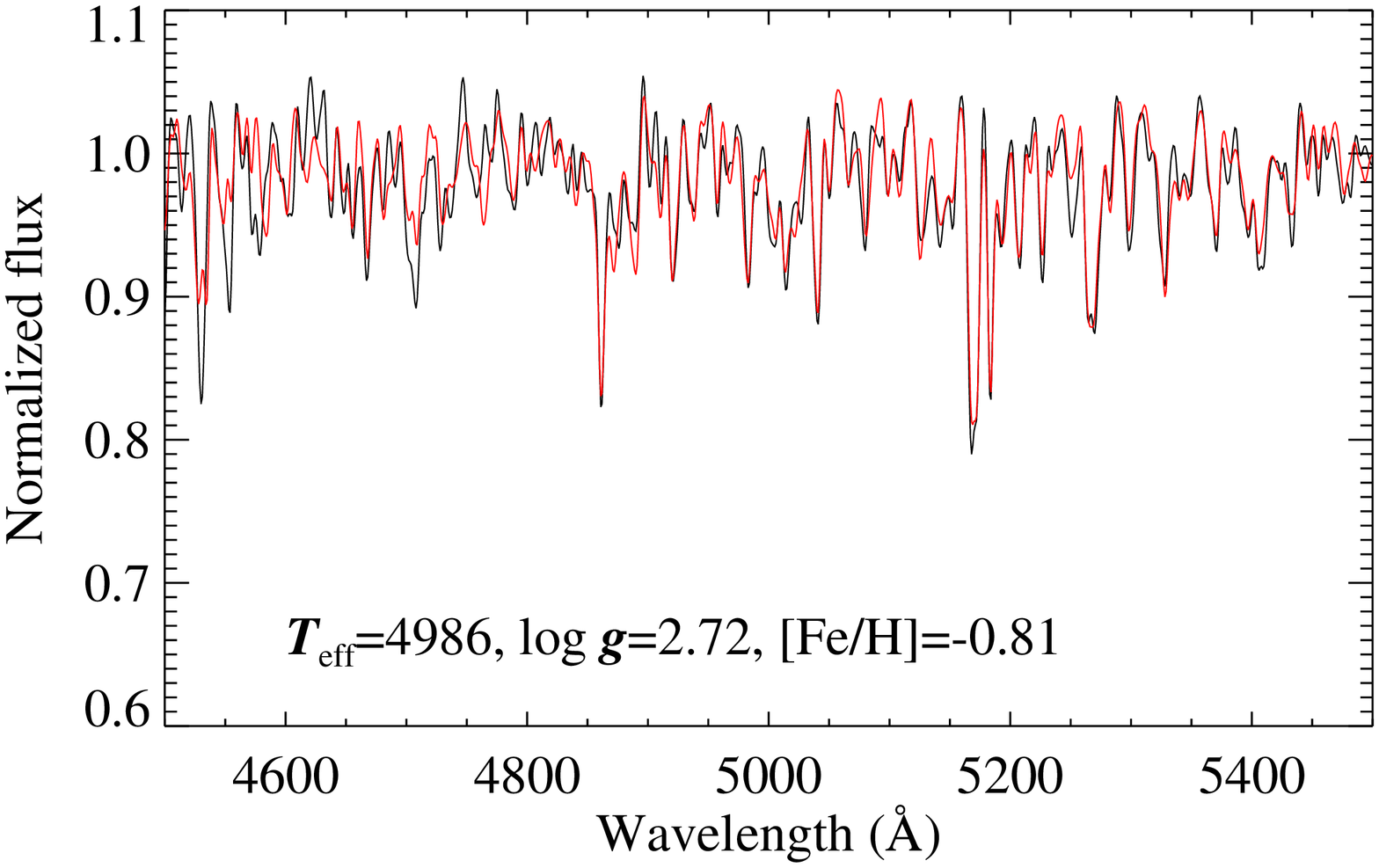}
\caption{Examples of the spectral matching of a synthetic spectrum. The black lines show the spectra (S/N$\sim70$, degraded to $R=1000$) of three cluster members (designated as star 29, 104, and 163, respectively, from the top to bottom panels). In each panel, the red line shows the best-fitting synthetic model spectrum, when the spectroscopic $\teff$ is forced to match the $(\jk)$-based IRFM scale.\label{fig:spectra}} \end{figure}

A few examples of the spectral matching with synthetic spectra are shown in Figure~\ref{fig:spectra}. In each panel, the black line is an observed spectrum with a relatively high S/N, and the red line represents the best-fitting synthetic spectrum. The best-fitting parameters shown in each panel are those derived when the spectroscopic $\teff$ is forced to match the $(\jk)$-based IRFM $\teff$ scale (see below). As an alternative to the above approach, we set [$\alpha$/Fe] in the models as a free parameter and derived [Fe/H] and  [$\alpha$/Fe] \citep[mostly from Mg~Ib triplet, see also][]{lee:11} along with other stellar parameters. For the $\wcen$ stars analyzed in this study, however, the net difference in [Fe/H] was small ($\Delta {\rm [Fe/H]}=0.05\pm0.01$~dex). The spectral matching was performed independently of distance or foreground reddening of the cluster.

\subsection{Systematic Errors in the Parameter Estimation}

\begin{figure}
\centering
\epsscale{1.05}
\plotone{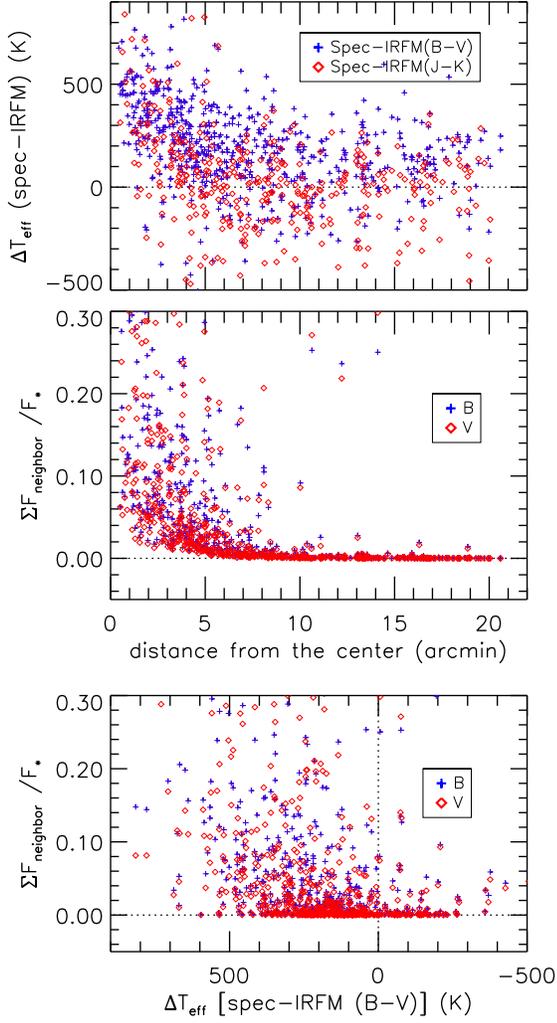}
\caption{Top: comparison between spectroscopically derived $\teff$ and IRFM $\teff$ as a function of a projected distance from the cluster's center. Differences are shown for $(\bv)$ (blue cross) and $(\jk)$-based (red diamond) IRFM relations, respectively. Middle: ratio of the summed flux from neighboring stars to a star's flux as a function of a distance from the cluster's center. The blue crosses and red diamonds represent flux ratios computed in the $B$ (blue cross) and $V$ (red diamond) passbands, respectively, assuming a Moffat profile with a FWHM of $2\arcsec$. Bottom: scatter plot of flux contamination by neighboring stars versus the difference between spectroscopic $\teff$ and $(\bv)$-based IRFM $\teff$. The symbols are the same as in the middle panel.\label{fig:teff}} \end{figure}

In the course of this study, we found a large difference between spectroscopic and photometric $\teff$ for stars in the central region of the cluster. Figure~\ref{fig:teff} shows comparisons between spectroscopic $\teff$ from the above spectral matching technique and IRFM $\teff$ based on $(\bv)$--$\teff$ (blue crosses) and $(\jk)$--$\teff$ (red diamonds) relations in \citet{gonzalez:09}. We took $BV$ and $JK_s$ photometry from RLR04 and the Point Source Catalog (PSC) of the Two Micron All Sky Survey \citep[2MASS;][]{skrutskie:06}, respectively. Only stars that have been observed more than once with S/N$\geq52$ are included in this comparison. We assumed $E(\bv)=0.11$ \citep{lub:02} for all stars, along with $R_V\equiv A_V/E(\bv)=3.1$, and $E(\jk)/E(\bv)=0.56+0.06\, (\jk)_0$ \citep{an:07}. We supplied [Fe/H] from our initial analysis based on the spectroscopically derived $\teff$ as an input to the IRFM relation.

As shown in the top panel of Figure~\ref{fig:teff}, differences between spectroscopic and photometric $\teff$ systematically increase toward the cluster's center. Both $\bv$ and $\jk$ IRFM relations essentially yield the same trend. The fact that the difference between spectroscopic and photometric $\teff$ remains fairly constant in the outer part of the cluster ($\ga 5\arcmin$ from the cluster's center) suggests that the strong $\teff$ disagreement in the central region is somehow related to the crowding of stars. Since $BVJK_s$ photometry in RLR04 and 2MASS was extracted based on modeling of a point spread function (PSF), we suspect that the problem lies in the contamination of spectroscopic fibers by a nearby star's flux in the crowded fields, given the relatively large aperture size of optical fibers used in this study ($\sim2\arcsec$ diameter on a focal plane).

To further investigate the above systematic difference in $\teff$, we estimated a fractional contribution of flux from neighboring stars using $BV$ photometry. For this experiment, we took RLR04 photometry and summed fluxes from all nearby stars within $10\arcsec$ from each star ($\Sigma F_{\rm neighbor}$). We assumed the Moffat profile of stellar light with $2\arcsec$ for a FWHM of the seeing disk. We divided $\Sigma F_{\rm neighbor}$ by a source flux at its centroid ($F_{\star}$) to estimate a fractional contribution of flux, $f_{\rm neighbor}$.\footnote{Practically, it is an inverse of a `separation' index in \citet{stetson:03} for computing the sum of flux from neighboring stars. We assumed that a stellar profile can be approximated by $(1+r^2)^{-2}$, where $r$ is a distance from the centroid in units of a characteristic length scale \citep[see also][]{clem:08}.} The results are shown in the middle panel of Figure~\ref{fig:teff} for the $B$ (blue pluses) and $V$ (red diamonds) passbands, respectively. The $f_{\rm neighbor}$ increases toward the cluster's center, implying that the degree of flux contamination by neighboring stars should also rise in the observed spectra. This is also shown in the bottom panel, where $f_{\rm neighbor}$ tends to be bigger when the difference between spectroscopic and photometric $\teff$ is larger. This suggests that the spectroscopic $\teff$ is likely overestimated for most of our spectra in the crowded inner region.

We simulated the effects of flux contamination by taking theoretical spectra with $\teff=5000$~K (representative of our sample stars) and combining them with those for fainter giants ($\teff=6000$~K) and hot horizontal branch stars ($\teff=10,000$~K) in the cluster. For a range of [Fe/H], we found that thelar $\teff$ from spectra can be overestimated by $\sim50$\,--\,$200$~K if the flux contamination is $10\%$ of the star's continuum at $4000$\,\AA. Spectroscopic $\teff$ can be overestimated by $\sim150$\,--\,$600$~K if the contamination increases to $30\%$. The $\logg$ and [Fe/H] are also driven to systematically larger values ($\Delta \logg \sim0.07$\,--\,$0.19$~dex and $\Delta {\rm [Fe/H]}\sim0.01$\,--\,$0.16$~dex for the $10\%$ contamination case), mainly because of the elevated $\teff$. However, more quantitative assessment of the bias requires full knowledge of the spectroscopic properties of contaminating sources and of the impacts of actual seeing conditions and source positions on the efficiency of fibers. Instead of correcting for these biases, we adopted photometric $\teff$ to minimize such biases, and searched for the best-fitting surface gravity and metallicity at fixed $\teff$.

Our adopted $\teff$ in the following analysis is based on the IRFM $\teff$ from $\bv$ colors [$\teff\, (\bv)$], but with a modification to put them on the $(\jk)$-based IRFM $\teff$ scale [$\teff\, (\jk)$]. We utilized $\bv$ data because they were extracted from images with a higher spatial resolution (FWHM$\sim1.0\arcsec$--$1.7\arcsec$; RLR04) than the 2MASS images (FWHM$\sim2.5\arcsec$--$2.7\arcsec$). However, $\teff\, (\bv)$ leads to systematically low [Fe/H] estimates for our sample stars compared to previous results in the literature (e.g., JP10). Therefore, we adjusted a zero-point of $\teff$ to be on the same scale as $\teff\, (\jk)$, which results in a better agreement of the MDF peak at [Fe/H]$\sim-1.75$ with that of JP10 (see below). We took photometry with $\sigma < 0.05$~mag in $\jk$ and $\sigma < 0.02$~mag in $\bv$, selecting only those lying outside of the crowded region ($r > 5\arcmin$ from the cluster's center), and derived $\teff\, (\jk)=\teff\, (\bv)+101.6-0.156\, [\teff\, (\bv)-5000]$. We used this relation to put $\teff\, (\bv)$ onto the $\teff\, (\jk)$ scale. Since IRFM relations depend on [Fe/H], a couple of iterations were required to finalize the stellar parameter estimates.

\begin{deluxetable*}{ccccccccccccc}
\tabletypesize{\scriptsize}
\tablewidth{0pt}
\tablecaption{Spectroscopic Properties of Stars in $\wcen$ \label{tab:tab1}}
\tablehead{
  \colhead{ID} &
  \colhead{R.A.\,(J2000.0)} &
  \colhead{Decl.\,(J2000.0)} &
  \colhead{$V$\tablenotemark{a}} &
  \colhead{\bv\tablenotemark{a}} &
  \colhead{$\teff$\tablenotemark{b}} &
  \colhead{$\logg$} &
  \colhead{[Fe/H]} &
  \colhead{[Ca/Fe]} &
  \colhead{$v_r$} &
  \colhead{$f_{\rm neighbor}$\tablenotemark{c}} &
  \colhead{$N_{\rm obs}$\tablenotemark{d}} &
  \colhead{Mem\tablenotemark{e}} \\
  \colhead{} &
  \colhead{(h:m:s)} &
  \colhead{($\arcdeg$:$\arcmin$:$\arcsec$)} &
  \colhead{(mag)} &
  \colhead{(mag)} &
  \colhead{(K)} &
  \colhead{(dex)} &
  \colhead{(dex)} &
  \colhead{(dex)} &
  \colhead{(km s$^{-1}$)} &
  \colhead{} &
  \colhead{} &
  \colhead{}
}
\startdata
1 & 13:26:34.43 &-47:30:24.4&14.401&1.101 &4793&$2.22\pm0.07$ &$-1.22\pm0.04$ & $0.24\pm0.01$ & $225.7\pm3.6$ & 0.05 &3& m \\
2 & 13:26:31.42 &-47:31:37.1&14.403&0.940 &5016&$1.75\pm0.41$ &$-1.78\pm0.07$ & $0.03\pm0.03$ & $214.8\pm8.0$ & 0.15 &3& m \\
3 & 13:26:22.49 &-47:23:54.6&14.404&0.902 &5045&$1.75\pm0.41$ &$-1.79\pm0.11$ & $0.53\pm0.04$ & $231.2\pm3.8$ & 0.01 &3& m \\
4 & 13:26:21.64 &-47:25:23.1&14.405&1.005 &4886&$2.10\pm0.15$ &$-1.65\pm0.10$ & $0.38\pm0.02$ & $235.3\pm6.0$ & 0.01 &3& m \\
5 & 13:26:57.96 &-47:26:23.4&14.406&0.920 &5021&$2.43\pm0.13$ &$-1.79\pm0.02$ & $0.30\pm0.02$ & $234.3\pm4.7$ & 0.02 &3& m \\
\enddata
\tablecomments{Only a portion of this table is shown here to demonstrate its form and content. A machine-readable version of the full table is available.}
\tablenotetext{a}{Photometry taken from RLR04.}
\tablenotetext{b}{Adjusted IRFM $\teff$ (see the text).}
\tablenotetext{c}{Fractional contribution of flux in the $B$ passband from neighboring stars.}
\tablenotetext{d}{Number of spectra with S/N$\geq52$.}
\tablenotetext{e}{Cluster membership based on radial velocities and proper motions: m -- cluster member. x -- non-member.}
\end{deluxetable*}

Our $\logg$ and [Fe/H] estimates from the above spectral matching and the photometric $\teff$ are tabulated in Table~\ref{tab:tab1}, along with a sequential star ID assigned in this study, celestial coordinates, a heliocentric radial velocity, a fractional contribution of flux from neighboring stars in the $B$ passband ($f_{\rm neighbor}$), and a number of good spectroscopic observations ($N_{\rm obs}$). Only those having at least one spectrum with S/N$\geq52$ are listed ($N=712$). Among them, $603$ stars have been observed more than once, for which we report averaged spectroscopic parameter estimates. Weights in this average were given by an S/N at the wavelengths used in the spectral matching. The errors represent the weighted standard deviations of these measurements. Median errors from stars with repeat measurements amount to $\sigma (\teff)=9$~K, $\sigma (\logg)=0.2$~dex, and $\sigma ({\rm [Fe/H]})=0.07$~dex. The derivation of [Ca/Fe] is described in the next section.

Among the high-resolution spectroscopic studies in the literature, \citet{marino:12} provided a reasonably large number of stars for comparison with our work. There are $20$ stars in common for a $1\arcsec$ search radius. For these stars, the mean difference is $\langle \Delta {\rm [Fe/H]}\rangle=0.08\pm0.03$~dex in the sense that their abundance is larger. They adopted thelar $\teff$ from the $V-K$ IRFM relation, which is cooler than our adjusted IRFM $\teff$ by $\langle \Delta {\rm \teff}\rangle=172\pm18$~K.

\subsection{[Ca/Fe]}\label{sec:cafe}

We estimated [Ca/Fe] based on measurements of \ion{Ca}{2} H \& K lines from our spectra. This part of the spectrum has a lower S/N ($\sim17$) than those used in the above spectral matching of synthetic spectra, but the lines are significantly stronger. We employed the Ca abundance indicator $A$(Ca) defined by \citet{norris:81}, but adopted continuum ranges in \citet{lim:15} to minimize the effect of CN absorption on the blue edge of the continuum: $A({\rm Ca)}=1-2 \bar{F}_{3916-3985}/(\bar{F}_{3894-3911}+\bar{F}_{3990-4025})$, where $\bar{F}_{\lambda_1-\lambda_2}\equiv\int^{\lambda_2}_{\lambda_1} F_{\lambda} d\lambda / (\lambda_2-\lambda_1)$ and $F_\lambda$ is the flux at $\lambda$. The spectra were normalized using a straight line, which was fit to the above continuum range, before estimating $A$(Ca).

\begin{figure}
\centering
\epsscale{1.05}
\plotone{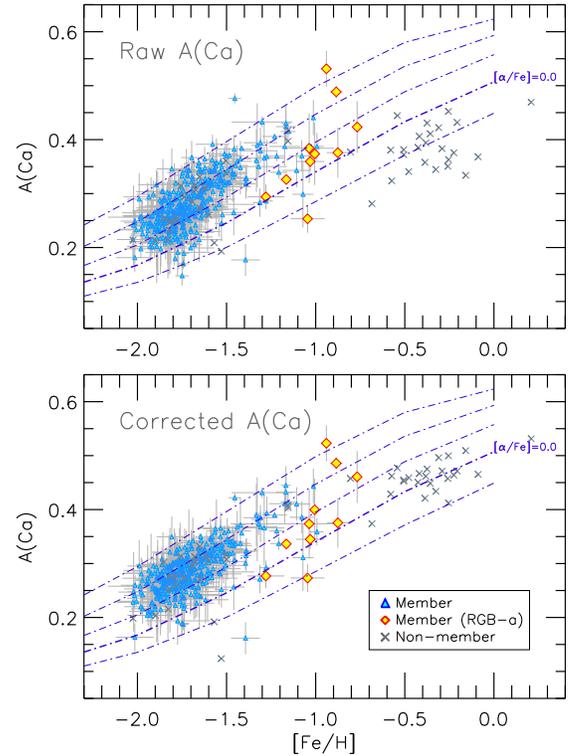}
\caption{\ion{Ca}{2} H \& K abundance indicator $A$(Ca) as a function of [Fe/H]. The top panel shows raw $A$(Ca) measurements, and the bottom panel displays reduced $A$(Ca) values to a reference point at $\teff=5000$~K and $\logg=2.0$. The dot-dashed lines represent theoretical predictions from PHOENIX models with varying [$\alpha$/Fe] from $-0.2$ to $0.6$ in steps of $0.2$~dex. The symbols are the same as in Figure~\ref{fig:cmd}, except for the gray cross for non-members.\label{fig:aca}} \end{figure}

Our $A$(Ca) measurements are displayed in the top panel of Figure~\ref{fig:aca} along with the [Fe/H] from the synthetic spectral matching. Only stars with $\sigma({\rm [Fe/H]})<0.2$~dex and $\sigma[A{\rm (Ca)}]<0.1$ from repeat measurements are shown. We excluded stars with a fractional contribution of flux from neighboring stars greater than $f_{\rm neighbor} =0.10$ (see the bottom panel in Figure~\ref{fig:teff}). The different symbols in Figure~\ref{fig:aca} are described below.

The \ion{Ca}{2} H \& K absorption line strength depends on $\teff$ and $\logg$. According to theoretical spectra from the PHOENIX model atmospheres \citep{husser:13}\footnote{Available at http://phoenix.astro.physik.uni-goettingen.de/}, hotter stars or stars with a higher $\logg$ have weaker absorptions at fixed metallicity by $\Delta A{\rm(Ca)}/\Delta \teff \approx 0.02/100$~K and $\Delta A{\rm(Ca)}/\Delta \logg \approx 0.02/0.5$~dex, respectively. These changes are large enough to influence our result on a [Ca/Fe] versus [Fe/H] diagram, since $\teff$ and $\logg$ are strongly tied to [Fe/H] for a magnitude-limited RGB sample.

We utilized synthetic spectra based on the PHOENIX models to reduce our measured $A$(Ca) values to a reference point at $\teff=5000$~K and $\logg=2.0$, and eventually to determine [Ca/Fe]. The synthetic spectra are available for a combination of [Fe/H] and [$\alpha$/Fe], where the alpha-abundance in their models includes O, Ne, Mg, Si, S, Ar, Ti, as well as Ca. We made a specific assumption that [$\alpha$/Fe] is equal to [Ca/Fe], and preceded to estimate theoretical A(Ca) values for a large grid of models that covers the parameter space of our sample stars. For the observed [Fe/H] and [Ca/Fe] of a star, we computed a difference between theoretical $A$(Ca) values at a star's physical parameters ($\teff$ and $\logg$) and the reference point, and then added the difference to the observed $A$(Ca) value. Even though we adopted ATLAS9 models in our determination of $\teff$, $\logg$, and [Fe/H], this differential correction in $A$(Ca) should be relatively free of a specific choice of model sets. The bottom panel of Figure~\ref{fig:aca} displays these corrected $A$(Ca) values.

The dot-dashed lines in Figure~\ref{fig:aca} represent theoretical predictions from the PHOENIX models at $\teff=5000$~K and $\logg=2.0$ with varying [$\alpha$/Fe] from $-0.2$ to $0.6$ in steps of $0.2$~dex. We derived the [Ca/Fe] of stars from their corrected $A$(Ca) based on these model lines. If the model spectra from ATLAS9 \citep{castelli:04} were used instead to convert the corrected $A$(Ca), the [Ca/Fe] values of stars would become systematically lower by $\sim0.1$~dex. This absolute scale change should be taken separately from the above differential corrections on $A$(Ca). Individual errors in [Ca/Fe] were computed as a quadrature sum of errors in $A$(Ca), $\teff$, $\logg$, and [Fe/H]. If there exist more than one spectrum, average [Ca/Fe] estimates and their weighted standard deviations are given in Table~\ref{tab:tab1}. The median error is $\sigma (\rm [Ca/Fe])=0.10$~dex from stars with repeat measurements.

\subsection{Cluster Membership}

We selected cluster members of $\wcen$ based on our radial velocity measurements, aided by the proper-motion membership probabilities in \citet{vanleeuwen:00}. The cluster has a large $v_r$ \citep[e.g.,][]{reijns:06}, and is well separated from foreground stars in kinematic space. For stars with a proper-motion membership probability greater than $70\%$, the mean of our $v_r$ measurements yields $233$\,km\,s$^{-1}$. We took these proper-motion members within $\pm3\sigma$ from the mean $v_r$ ($192$\,km\,s$^{-1} \leq v_r \leq 275$\,km\,s$^{-1}$) as a member of $\omega$ Cen.

Among $712$ stars that have good-quality spectra (S/N$\geq52$), $581$ stars passed the above selection criteria. In the following analysis, we focus on a subset of these members ($N=410$) with good spectroscopic parameter estimates from multiple measurements: $\sigma({\rm [Fe/H]})<0.2$~dex, $\sigma[A{\rm (Ca)}]<0.1$, and $f_{\rm neighbor} <0.10$. These kinematically selected cluster members are shown by the blue triangles in Figures~\ref{fig:cmd} and \ref{fig:aca}. Among these, RGB-a members are highlighted by red diamonds. In Figure~\ref{fig:aca}, foreground stars are marked by a gray cross, most of which have the solar [$\alpha$/Fe] with high [Fe/H].

\section{Results}\label{sec:result}

\subsection{Metallicity Distribution Function}

\begin{figure}
\centering
\epsscale{1.05}
\plotone{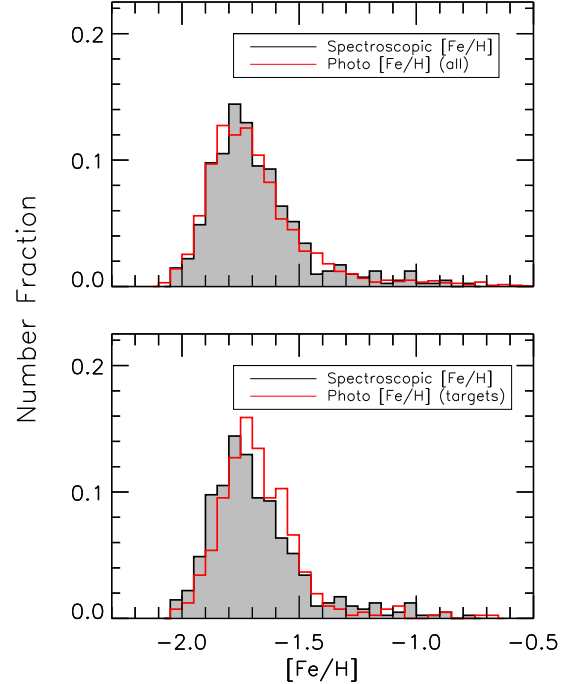}
\caption{Comparisons of the spectroscopic MDF (gray shaded histogram) with photometric MDFs (red histogram). Top: photometric MDF from all cataloged objects in RLR04 with the same color and magnitude ranges as in our sample. Bottom: photometric MDF for the spectroscopic targets only.\label{fig:mdf}} \end{figure}

In Figure~\ref{fig:mdf}, an MDF of our sample stars is shown by a gray shaded histogram. Only stars with good spectroscopic parameter estimates (blue triangles and red diamonds in Figure~\ref{fig:aca}) are included in the histogram. Our spectroscopic MDF is compared to two different photometric MDFs, both of which were constructed using the $BV$ data in RLR04. To compute photometric metallicities, we derived a fiducial line along the observed RGB on the $\bv$ versus $V$ CMD by fitting a line to a subset of our sample stars with $-1.9 < {\rm [Fe/H]} < -1.7$. At a given $V$, a displacement in $\bv$ from this fiducial line shows a tight correlation with spectroscopic [Fe/H]. We used a second-order polynomial to relate the $\bv$ displacement with [Fe/H].

In the top panel of Figure~\ref{fig:mdf}, a photometric MDF is shown for all objects listed in RLR04 that satisfy our color and magnitude sample cut. In the bottom panel, a photometric MDF for our spectroscopic sample is compared to the spectroscopic MDF. Both photometric MDFs were convolved with a gaussian error of $\sigma=0.07$~dex to provide an approximate match to the width of the main peak at [Fe/H]$\sim-1.75$ in the spectroscopic MDF. A small-scale difference in [Fe/H] of the order of $0.05$~dex is seen in the bottom panel. This was caused by a relatively steep $\bv$ versus [Fe/H] relation for the metal-poor stars, which was not fully accounted for with a simple second-order polynomial. Nevertheless, the difference is sufficiently small enough to draw the conclusion that the overall shapes of spectroscopic and photometric MDFs agree well. This suggests that our sample stars were randomly drawn from the photometric catalog and that our spectroscopic sample is a good representation of true underlying RGB populations in the cluster.

\begin{figure}
\centering
\epsscale{1.05}
\plotone{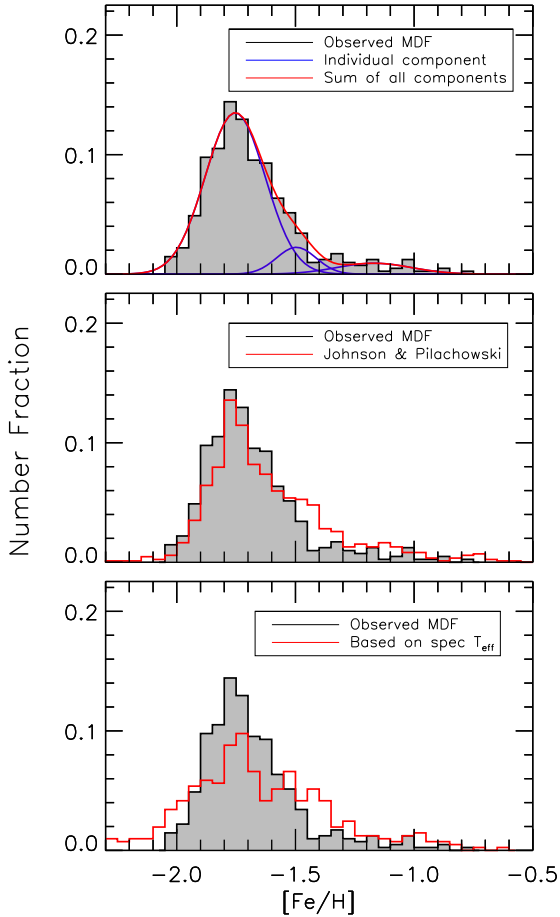}
\caption{Top: decomposition of the spectroscopic MDF (gray shaded histogram). Individual components of gaussian curves are shown by the blue lines, and the sum of these is drawn by a red line. Middle: comparison of the spectroscopic MDF with that of JP10 (red histogram). Bottom: same as in the middle panel, but with a comparison to an MDF constructed using spectroscopically derived $\teff$ (red histogram).\label{fig:mdf2}} \end{figure}

In the top panel of Figure~\ref{fig:mdf2}, we decomposed our spectroscopic MDF with three gaussian functions, as shown by blue solid lines. Each of the gaussian functions was fit at [Fe/H]$<-1.6$, $-2.6<{\rm [Fe/H]}<-1.3$, and [Fe/H]$>-1.6$ in sequential order, after removing a contribution from a preceding fit. The red line is a sum of these components. The gaussian peaks are located at [Fe/H]$=-1.75$, $-1.50$, and $-1.17$, respectively, each of which corresponds to RGB-MP, RGB-MInt1, and a combination of RGB-MInt2+3 and RGB-a in JP10. The fractional contribution of RGB-MP to the total sample is approximately $85\%$, while that of RGB-MInt1 is $\sim10\%$.

In comparison to an MDF in JP10, which is shown by a red histogram in the middle panel of Figure~\ref{fig:mdf2}, fewer stars are observed at [Fe/H]$\ga-1.5$ in our MDF, while a relatively larger number of metal-poor stars are seen at [Fe/H]$\la-1.5$. JP10 estimated $61\%$ and $27\%$ for a fraction of RGB-MP and RGB-MInt1, respectively. The null hypothesis that the two samples from this study and JP10 were drawn from the same underlying population is rejected at the $99.999 \%$ level of significance ($p=0.00001$) in the two-sided Kolmogorov--Smirnov (K-S) test. The narrower metal-poor peak at [Fe/H]$\sim-1.75$ in JP10 may indicate more accurate metallicity measurements in their analysis. However, a simple convolution of their measurements, in order to match the width of the metal-poor peak in our MDF, would not simply reduce the relative number of intermediate metallicity populations at [Fe/H]$\sim-1.5$. JP10 demonstrated that their MDF is similar in shape with those in \citet{norris:96} and \citet{suntzeff:96}, suggesting that our MDF also disagrees with these studies.

The bottom panel in Figure~\ref{fig:mdf2} shows a comparison of our spectroscopic MDF with an MDF from [Fe/H] estimates based on spectroscopically derived $\teff$ (red histogram). As discussed in \S~\ref{sec:param}, we adopted IRFM $\teff$ in the spectroscopic analysis because of a strong departure of spectroscopic $\teff$ from the fundamental photometric $\teff$ scale, most likely due to flux contamination of spectra by neighboring stars. On average, spectroscopic $\teff$ values are higher than the IRFM estimates, and are stretched toward higher values for more metal-rich stars. Since a higher $\teff$ generally leads to a higher [Fe/H] in our spectral matching technique, relatively more metal-rich stars are overrepresented in the `pure' spectroscopic MDF (red histogram), displaying a skewed distribution toward higher [Fe/H]. The frequency of its intermediate metallicity population seems even higher than that in JP10.

\subsection{Radial Metallicity Gradient}

A number of studies argued for a weak radial metallicity gradient in $\wcen$, with metal-rich stars being more centrally concentrated than metal-poor stars in the cluster \citep[e.g.,][]{norris:96,suntzeff:96,bellini:09,johnson:10}. There has also been an indication that metal-rich stars in the cluster tend to have a smaller velocity dispersion \citep{norris:97,sollima:05}.

\begin{figure}
\centering
\epsscale{1.0}
\plotone{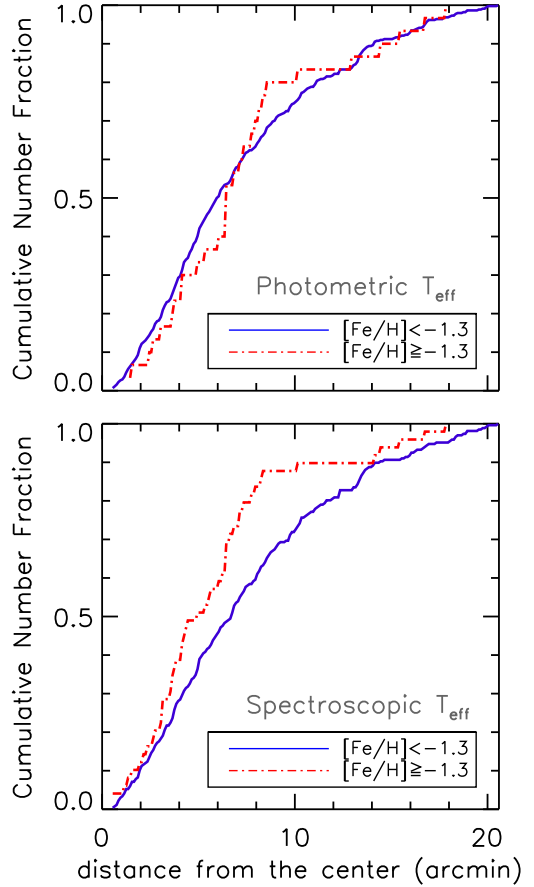}
\caption{Top: comparison of cumulative radial distributions of stars with [Fe/H]$<-1.3$ (blue solid line) and [Fe/H]$\geq-1.3$ (red dot-dashed line), based on the [Fe/H] computed using photometric $\teff$. Bottom: same as in the top panel, but based on spectroscopically derived $\teff$.\label{fig:radial}} \end{figure}

We divided our sample stars into two groups based on [Fe/H], and compared their cumulative distributions with each other. Such comparisons are shown in the top panel of Figure~\ref{fig:radial} for stars with [Fe/H]$\geq-1.3$ and [Fe/H]$<-1.3$, respectively, where we took [Fe/H] estimates based on the IRFM $\teff$. Their radial distributions are similar to each other, and a hypothesis that they are drawn from the same underlying distribution cannot be rejected in a two-sided K-S test ($p=0.64$). The above division on [Fe/H] was made to separate (relatively) metal-rich stars from the bulk of the RGB-MP population, but a range of the division ($-1.6 \leq {\rm [Fe/H]} \leq -1.2$) resulted in a similar statistical significance ($0.24 < p < 0.68$).

On the other hand, a completely different picture emerges on the radial metallicity gradient if spectroscopic $\teff$ is adopted. This is shown in the bottom panel of Figure~\ref{fig:radial}, in which metal-rich stars ([Fe/H]$\geq-1.3$) are more centrally concentrated than metal-poor stars.  According to the K-S test, the two groups have significantly different spatial distributions ($p=0.010$). The result is even more significant if the division was made at [Fe/H]$=-1.5$ or $-1.6$, yielding $p<0.001$. However, as discussed above, our spectroscopic $\teff$ is likely affected by the blending of light in a crowded field, which results in an overproduction of more metal-rich stars (see the bottom panel in Figure~\ref{fig:mdf2}). Consequently, more metal-rich stars are biased toward the cluster's center, as shown in the bottom panel of Figure~\ref{fig:radial}. In this example, we did not apply a cut based on $f_{\rm neighbor}$ in order to demonstrate the impact of crowding in the spectroscopic analysis, but the statistical significance remains the same even if $f_{\rm neighbor}<0.10$ is set.

Figure~\ref{fig:radial} demonstrates that blending of stellar light in the crowded central region of a cluster can easily lead to an apparent radial metallicity gradient if spectroscopically derived $\teff$ values are taken in the estimation of [Fe/H] without any necessary corrections. Since photometric $\teff$ makes [Fe/H] estimates less affected by such bias, we contend that our data do not support a radial metallicity gradient in $\wcen$. Nevertheless, it is noted that some previous spectroscopic studies based on photometric $\teff$ \citep[e.g.,][]{norris:96,johnson:10} suggested the presence of a weak metallicity gradient in the cluster. Their results are in line with the photometric metallicity gradient found by \citet{bellini:09}, although there was no convincing evidence for the metallicity gradient from the photometry in RLR04. More sophisticated approaches are needed to evaluate the impact of crowding on the spectroscopic metallicity estimation.

\subsection{[Ca/Fe] vs. [Fe/H]}

\begin{figure*}
\centering
\epsscale{0.85}
\plotone{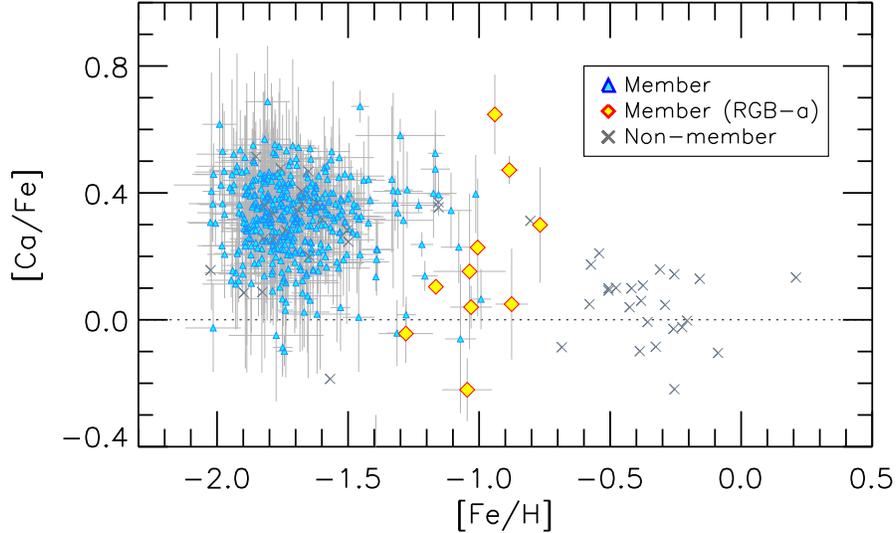}
\caption{[Ca/Fe] versus [Fe/H] diagram. The [Ca/Fe] estimates are based on the corrected $A$(Ca) values. The symbols are the same as in Figure~\ref{fig:aca}.\label{fig:cafe}} \end{figure*}

In Figure~\ref{fig:cafe}, [Ca/Fe] estimates\footnote{As described in \S~\ref{sec:cafe}, our [Ca/Fe] abundance ratios were estimated from a comparison of the \ion{Ca}{2} H \& K line indicator, $A$(Ca), with synthetic model spectra. We remind the reader that we took [$\alpha$/Fe] in these models as being equal to [Ca/Fe].} are shown for stars with good spectroscopic parameter estimates as a function of [Fe/H]. RGB-a members are highlighted by red diamonds. Foreground stars are indicated by a gray cross. While calcium abundance has been extensively used in the literature to study the chemical diversity of $\wcen$ stars \citep[e.g.,][]{norris:96}, it has been adopted mainly as a proxy of the bulk abundance of a star in such studies. JP10 provided the most recent estimates of both [Fe/H] and [Ca/Fe] for a large number of stars. The bulk median [Ca/Fe] of our sample is $0.33$~dex, which is close to $\langle {\rm [Ca/Fe]} \rangle = 0.29$ in JP10.

\begin{figure}
\centering
\epsscale{1.05}
\plotone{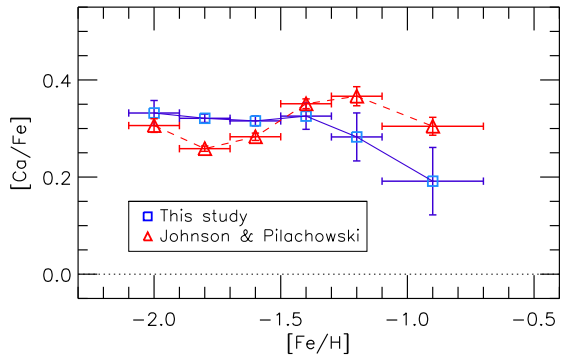}
\caption{Mean [Ca/Fe] as a function of [Fe/H] for $\wcen$ members (blue box). The mean [Ca/Fe] was computed in bins of $\Delta {\rm [Fe/H]}=0.2$--$0.4$~dex, and is connected by a blue solid line. The mean values for the JP10 sample are shown by a red triangle, and are connected by a red dashed line. The error bars indicate an error in the mean [Ca/Fe] in each [Fe/H] bin.\label{fig:cafe2}} \end{figure}

Unlike what has been suggested by JP10 \citep[see also][]{norris:95}, our sample essentially exhibits a flat [Ca/Fe] as a function of [Fe/H]. In Figure~\ref{fig:cafe2}, the blue boxes display our mean [Ca/Fe] values as a function of [Fe/H]. We computed the mean [Ca/Fe] in bins of $\Delta {\rm [Fe/H]}=0.2$--$0.4$~dex, taking the same set of $\wcen$ members as in Figure~\ref{fig:cafe}. The error bars indicate an error in the mean [Ca/Fe]. The binned [Ca/Fe] values are fairly flat for the metal-poor and metal-intermediate stars in the cluster. The same binning was applied to the JP10 sample, and their mean [Ca/Fe] values are shown by red triangles. As noted in JP10, their [Ca/Fe] estimates rise at [Fe/H]$\ga-1.5$ almost by $\Delta {\rm [Ca/Fe]}\sim0.1$~dex.

\begin{figure}
\centering
\epsscale{1.05}
\plotone{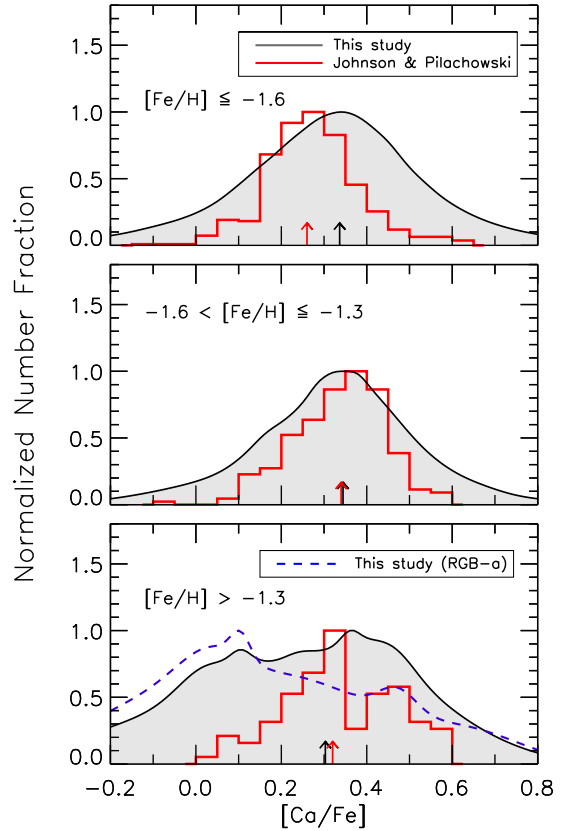}
\caption{Generalized histograms of [Ca/Fe] in three different [Fe/H] bins (gray shaded region). The red solid histogram represents [Ca/Fe] distributions from the JP10 sample. Each of the histograms is normalized to its maximum value. In each panel, the median [Ca/Fe] values are indicated by the black (this study) and red (JP10) arrows, respectively. In the bottom panel, a generalized histogram for the RGB-a stars from this work is additionally shown by the blue dashed line.\label{fig:cahist}} \end{figure}

This is also demonstrated in Figure~\ref{fig:cahist} for the three [Fe/H] bins with [Fe/H]$\leq-1.6$ (top), $-1.6<{\rm [Fe/H]}\leq-1.3$ (middle), and [Fe/H]$>-1.3$ (bottom), each of which encompasses RGB-MP, RGB-MInt1, and a combination of RGB-MInt2+3 and RGB-a, respectively. In each bin, we constructed a generalized histogram (gray shaded region), assuming a normal distribution of errors in our [Ca/Fe] estimates. The red solid histograms show [Ca/Fe] distributions from JP10. In the latter case, the mean [Ca/Fe] for the RGB-MP population is $\sim0.1$~dex lower than that of RGB-MInt1 (see the red upper arrow indicating their median [Ca/Fe]). JP10 found that both calcium and silicon abundances ([Si/Fe]) rise as [Fe/H] increases, and suggested mass-dependent or metallicity-dependent SN~II yields in the progenitor dwarf galaxy as a possible mechanism. However, our data with a constant [Ca/Fe] indicate that such scenario may not (or at least to a lesser degree) be necessary.

In addition, JP10 noted a hint of a bimodal distribution of [Ca/Fe] at $-1.3 < {\rm [Fe/H]} < -0.9$, with peaks at [Ca/Fe]$=+0.25$ and $+0.45$, in approximately equal proportions of stars. Their [Ca/Fe] distribution for RGB-MInt2+3 and RGB-a is reproduced in the bottom panel of Figure~\ref{fig:cahist}, which also exhibits their claimed double peaks. From $25$ stars with [Fe/H]$>-1.3$ in our sample, however, we could not verify the bimodal distribution of [Ca/Fe] at their proposed locations. The observed distribution is broad, covering approximately solar to [Ca/Fe]$\sim+0.7$. It is less likely that the dual-peak signatures are erased by large errors in our estimates, since the observed width of the [Ca/Fe] distribution is comparable to those in JP10 for more metal-poor stars.\footnote{Generalized histograms look broader than raw distributions because of convolution with errors.}

Similarly, we found a significantly larger scatter of [Ca/Fe] for RGB-a stars than seen in JP10. There are ten RGB-a stars included in our sample (red diamonds in Figure~\ref{fig:cafe}). Their mean [Ca/Fe] is $+0.17$~dex with a standard deviation of $\sigma=0.25$~dex. On the other hand, JP10 found a narrower distribution of RGB-a stars, with $\sigma=0.13$~dex (and a mean of $+0.26$~dex from $21$ stars). Including $3$ more stars in our sample that only have a single spectrum (with S/N$\geq52$) does not add much to the discussion.

Although we could not identify double peaks in [Ca/Fe], as proposed by JP10, there may be two underlying [Ca/Fe] distributions in other locations. As shown in the bottom panel of Figure~\ref{fig:cahist}, there appears to be a secondary peak at [Ca/Fe]$\approx+0.1$, in addition to a peak at [Ca/Fe]$\approx+0.35$. The latter seems to be an extension of the distribution from metal-poor members. On the other hand, the generalized histogram for RGB-a members (blue dashed line) indicates that the secondary peak is produced by the contribution from the most metal-rich stars in the cluster. This may even suggest a downturn of [Ca/Fe] for some of the stars in $\wcen$ \citep[e.g.,][]{pancino:02,origlia:03}. However, the scatter of our [Ca/Fe] estimates in the metal-rich bins is too large, with no well-defined peak, to draw any firm conclusions, such as the degree of Type~Ia contributions to the chemical evolutions of surviving $\wcen$ members.

\section{Summary and Discussion}

We have obtained and analyzed medium-resolution spectra for more than $700$ stars in the field of $\wcen$, and presented their [Fe/H] and [Ca/Fe]. For $\sim400$ kinematic members in the cluster's RGB with best spectroscopic parameter estimates, we found smaller fractions of relatively metal-rich stars in the cluster compared to those found in JP10. The fractional contribution of the metal-intermediate population peaked at [Fe/H]$\sim-1.5$ (RGB-MInt1) is approximately $10\%$, which is almost a factor of three lower than that in JP10. Our calcium abundance measurements based on \ion{Ca}{2} H \& K lines yield essentially a constant mean [Ca/Fe] at [Fe/H]$<-1.3$. More metal-rich stars, including the most metal-rich population in the cluster (RGB-a), show a large scatter in [Ca/Fe]. Our data do not support a bimodal [Ca/Fe] distribution in the previously claimed locations, although the number of metal-rich stars ([Fe/H]$>-1.3$) in our sample is small. Additional work is needed to increase the sample size with good spectroscopic estimates.

Both JP10 and this study relied on observations using the Hydra at the CTIO 4 m telescope. As we discussed in \S~\ref{sec:param}, one of the main caveats of optical spectroscopy in crowded regions is that neighboring stars can contribute fluxes to the observed spectra of targets, such as in the core of $\wcen$. Images of crowded fields can be processed through PSF-based fitting tools to separate flux from nearby stars and extract cleaned photometry of a target. Unlike in photometry, however, spectroscopic data are irreducible, in the sense that the degree of flux contamination from neighboring stars depends on seeing conditions and the actual positions of spectroscopic fibers, which are often difficult to keep track of. More careful absorption line analysis of blended spectra also requires accurate spectral information of neighboring stars in various evolutionary stages.

For our sample stars, we found that crowding generally raises spectroscopic temperatures due to enhanced continuum levels and H$\beta$ absorption line strengths by nearby hot stars. A higher $\teff$ leads to a higher [Fe/H], causing a skewed MDF. It also makes [Ca/Fe] estimates systematically lower. The net result is an apparent increase of more metal-rich and (relatively) Ca-weak stars in the cluster's central region. To relieve the effects of crowding, we adopted IRFM $\teff$ in this study, and removed stars that may have been severely contaminated by nearby objects. The latter procedure is a simple remedy, however, and is not an essential way of correcting for fundamental biases inherent in the data.

The effect of crowding is perhaps less severe in the analysis of JP10, because their sample is composed of the brightest giants in the cluster. As in this study, JP10 also adopted IRFM $\teff$ using $V\, -\,K_s$ in their spectroscopic analysis. They took $V$ magnitudes from heterogeneous photometric data in the literature compiled by \citet{vanleeuwen:00}, and combined them with 2MASS $K_s$. For stars with the same brightness as in our sample, we found that the $V$-band magnitudes in \citeauthor{vanleeuwen:00} tend to become systematically brighter toward the cluster's center than those in RLR04. A mean difference is $\langle \Delta V \rangle=0.102\pm0.003$ for stars within $r<5\arcmin$ from the cluster's center, while a difference is small ($\langle \Delta V \rangle=0.025\pm0.003$) in the outer region ($10\arcmin<r<20\arcmin$). A bluer $V\, -\, K_s$ color by $0.1$~mag corresponds to a $\sim70$~K increase in $\teff$, which typically makes an \ion{Fe}{1} abundance estimate higher by $\sim0.1$~dex. This systematic change can lead to an apparent metallicity gradient in the cluster, and can also make a MDF skewed toward a higher metallicity. The spatially varying photometric zero-point error could be responsible for the observed difference in MDF between JP10 and this study, but its impact on spectroscopic parameter estimates should be quantitatively assessed, which is beyond the scope of this paper.

\acknowledgements

We thank the referee for helpful comments and suggestions. D.A.\ and J.I.J.\ thank Dongwook Lim for useful discussions on calcium index measurements. Support for this work was provided by the National Research Foundation (NRF) of Korea to the Center for Galaxy Evolution Research (No.\ 2010-0027910). D.A.\ and J.I.J.\ acknowledge partial support provided by the Basic Science Research Program through the NRF of Korea funded by the Ministry of Education (NRF-2015R1D1A1A09058700). This work was developed from a master's thesis conducted by J.I.J.\ under the supervision of D.A. Y.S.L.\ acknowledges support provided by the Basic Science Research Program through the NRF of Korea funded by the Ministry of Science, ICT \& Future Planning (NRF-2015R1C1A1A02036658). S.C.R.\ acknowledges support provided by the Basic Science Research Program through the NRF of Korea funded by the Ministry of Education, Science, and Technology (NRF-2015R1A2A2A01006828).

This work is based on observations at the Cerro Tololo Inter-American Observatory, National Optical Astronomy Observatory (NOAO Prop. ID: 2005A-0166; PI: S.\ C.\ Rey), which is operated by the Association of Universities for Research in Astronomy (AURA) under a cooperative agreement with the National Science Foundation. 

\facility{Blanco (Hydra)}

{}

\end{document}